\documentclass[preprint,showpacs,preprintnumbers,amsmath,amssymb]{revtex4}

\usepackage{graphicx}
\usepackage{dcolumn}
\usepackage{bm}

\begin{document}


\title{Which group velocity of light in a dispersive medium?}

\author{Omar El Gawhary}
\affiliation{Optics Research Group, Department of Imaging Science and Technology, Delft University of Technology,\\
Lorentzweg 1, 2628 CJ Delft, The Netherlands}%

\author{Sergio Severini}
\affiliation{Centro Interforze Studi per le Applicazioni Militari\\
Via della Bigattiera 10, 56122 San Piero a Grado (Pi), Italy}%

\author{Paolo Christillin}
\affiliation{Dipartimento di Fisica, Universit$\grave{a}$ di Pisa\\
Largo Bruno Pontecorvo 3, 56127 Pisa, Italy}

\date{\today}

\begin{abstract}
The interaction between a light pulse, traveling in air, and a
generic linear, non-absorbing and dispersive structure is analyzed. It is shown that energy conservation imposes a constraint between the group velocities of the transmitted and reflected light pulses. It follows that the two fields propagate with group velocities depending on the dispersive properties of the environment (air) and on the transmission properties of the optical structure, and are one faster and the other slower than the incident field. In other words, the group velocity of a light pulse in a dispersive medium is reminiscent of previous interactions.
One example is discussed in detail.
\end{abstract}

\maketitle

\begin{figure}
    \centering
        \includegraphics[width=0.5\textwidth]{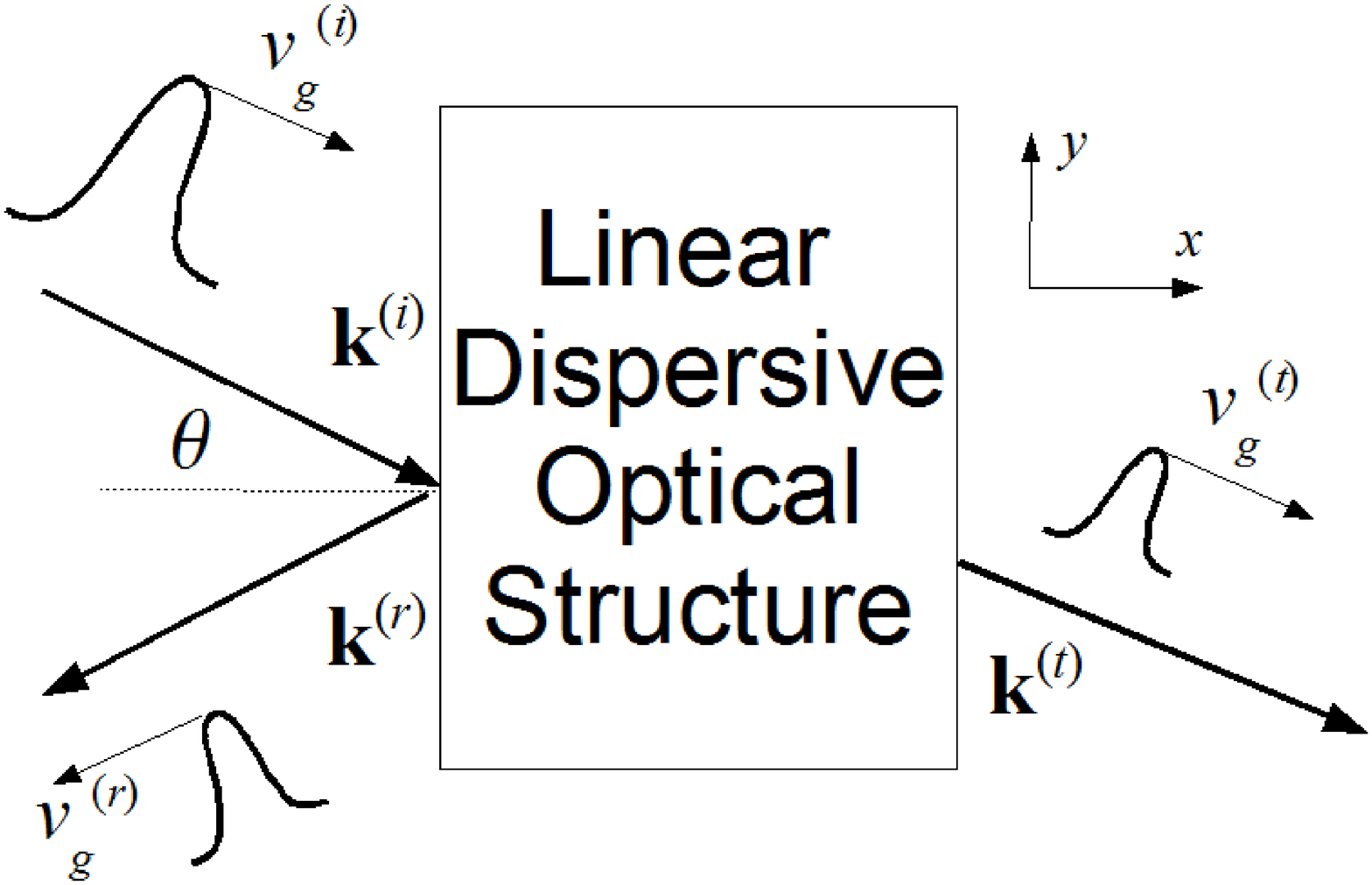}
       \caption{An optical beam incident over
       a linear dispersive and finite optical structure
       (such as a Beam Splitter or a PBG or a Lens etc.) is represented.
       The transmitted and reflected beams are schematically
       depicted as two other smaller wave packets, traveling in different directions.}
       \label{interazione}
\end{figure}

\begin{figure}
    \centering
        \includegraphics[width=0.48\textwidth]{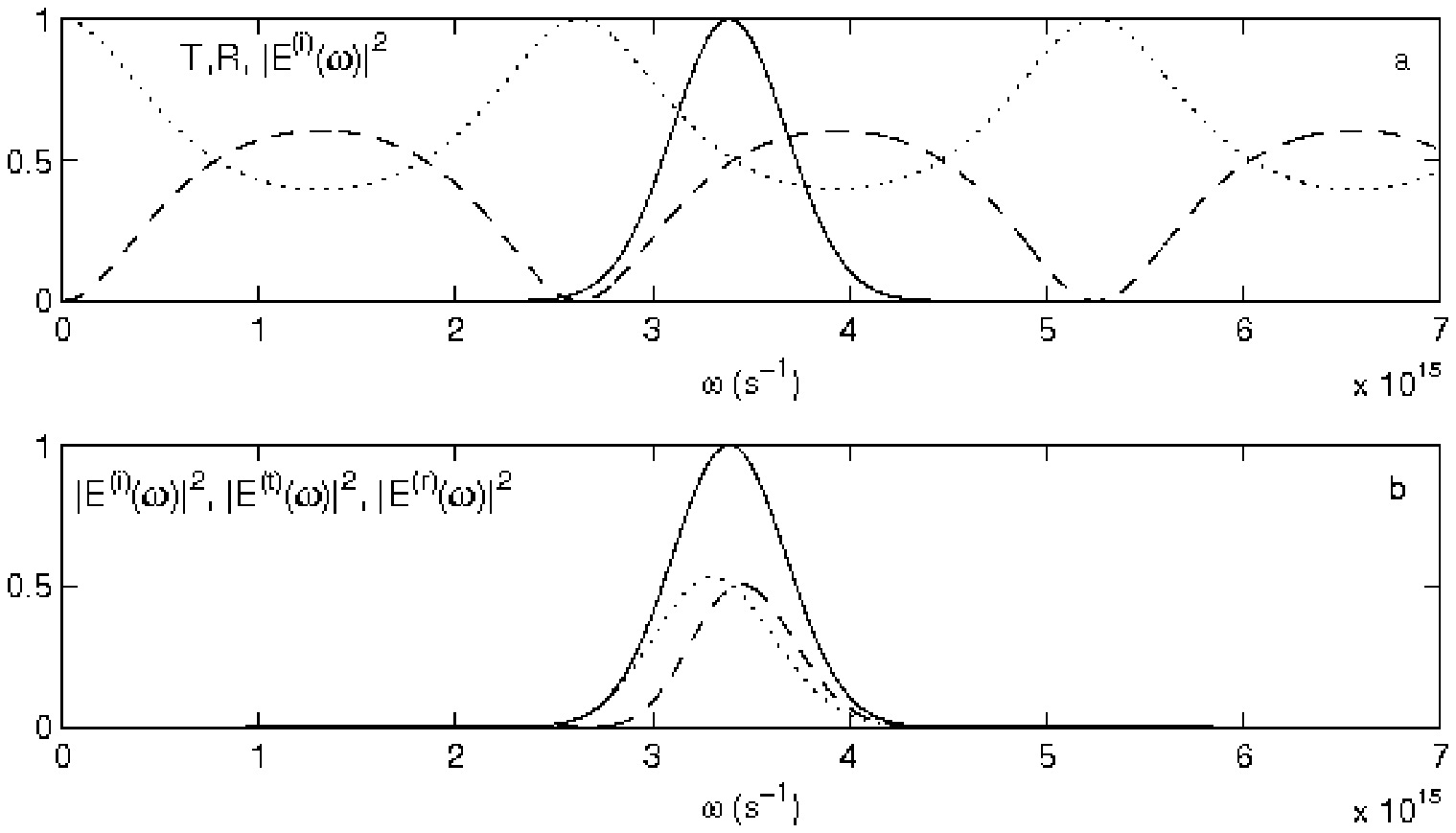}
        \caption{Transmitted (dotted line) and reflected (dashed line) field spectra originating from a rectangularly symmetric slab of  thickness $0.3\mu m$ and refractive index $n=1,5$. The incident field (continuous line) has an angle of incidence $\theta_i=65^\circ$.}
        \label{spettri}
\end{figure}

The concept of group velocity of an electromagnetic wave packet,
born in the early $19^{th}$ century \cite{Brillouin}, has been
thoroughly analyzed in many propagation regimes: for dispersive,
non-dispersive and absorbing properties of the propagation
environment \cite{Puri}. Such a quantity is naturally linked to the
propagation velocity of energy. In fact, if the optical field
propagates in dispersive and non-absorbing media, the two concepts
of group and energy velocity tend to coincide \cite{Brillouin}. In
recent years this topic has induced a good deal of studies mainly
within the context of the superluminal behavior of light, where the
role of several phenomena, like material gain, Kerr nonlinearities,
birifrangence, has been examined in detail \cite{Govyadinov06}.
Apart from all these interesting results, in our opinion there still
exists some simple, but intriguing, aspect in this scenario that
deserves to be studied. Actually, in the present Letter we consider
a wave packet incident over a generic dispersive, non-absorbing
optical structure like a dielectric slab, a Beam Splitter or a
finite Photonic Band Gap (PBG) etc.. On general grounds, due to the
interaction of the wave packet with this optical structure, we have
two emerging fields: the transmitted and the reflected pulse (as
schematically depicted in Fig.\ref{interazione}). We tackle the
problem of their group velocities. As above mentioned, we will only
assume that the propagation environment is dispersive, like air is,
and that the involved field frequencies are far from the absorbtion
lines. It will be shown that, due to the combination of the
principle of energy conservation, the dispersive properties of the
propagation environment and the transmission properties of the
generic optical structure, {\it differences among the group
velocities of incident, transmitted and reflected fields arise}. For
the sake of simplicity, and without loss of generality, we will
refer only to fields having the wave vector ${\bf k}$ lying in the
$x,y$ plane, as shown in Fig.\ref{interazione}. We can expand the
electric field $E^{(i)}(x,y,t)$ of the incident wave packet, in
plane waves:
\begin{equation}\label{campo incidente}
E^{(i)}(x,y,t)=\frac{1}{2\pi}\int E^{(i)}(\omega)
\exp[i({\bf k}^{(i)}\cdot {\bf r}-\omega t)]d\omega
\end{equation}
where ${\bf r}=(x,y)$. In Eq.\ref{campo incidente}  ${\bf
k}^{(i)}=k(\omega){\bf s}^{(i)}$, where ${\bf s}^{(i)}=( \cos\theta,
- \sin\theta)$ and $\theta$ is the angle of incidence, and
\begin{equation}\label{k}
k(\omega)=n(\omega)\omega/c.
\end{equation}
In Eq.\ref{k}, $n(\omega)$ is the refractive index of the
environment which in nonabsorbing media is a real function of a real
variable. As usual, the input-output behavior of a generic
structure, with respect to the incident field, is describable by
means of reflection and transmission coefficients, $r(\omega)$ and
$t(\omega)$ respectively, connected with the (intensity)
reflectivity and transmissivity through the well known relationships
$R(\omega)=\left|r(\omega)\right|^2$ and
$T(\omega)=\left|t(\omega)\right|^2$. Under such hypotheses
\cite{BornWolf}, \cite{Fowles75}, \cite{Barnett}
\begin{equation}\label{conserv energia}
T(\omega)+R(\omega)=|t(\omega)|^2+|r(\omega)|^2=1.
\end{equation}
The transmitted and reflected beams are
\begin{eqnarray}\label{campoTx}
E^{(t)}(x,y,t) & =& \frac{1}{2\pi}\int t(\omega)E^{(i)}(\omega)\nonumber\\
 & \times & \exp [i{\bf k}^{(t)}\cdot {\bf r}-\omega t)] d\omega  \\ \label{camporx}
E^{(r)}(x,y,t) & = & \frac{1}{2\pi}\int r(\omega)E^{(i)}(\omega)\nonumber\\
& \times & \exp [i({\bf k}^{(r)}\cdot {\bf r}+\omega t )]d\omega
\end{eqnarray}
where ${\bf k}^{(r)}= k(\omega) {\bf s}^{(r)}$, ${\bf k}^{(t)}=
k(\omega) {\bf s}^{(t)}$ and ${\bf s}^{(r)}=( -\cos\theta, -
\sin{\theta})$ and ${\bf s}^{(t)}=( \cos\theta, - \sin{\theta})$.
Group velocities are defined as
\begin{equation}\label{gruppo_incidente}
\mathbf{v_g}^{(\alpha)}=\left(\frac{d\omega}{dk}\right)_{k_0^{(\alpha)}}
\mathbf{s}^{(\alpha)}=\left(\frac{dk}{d\omega}\right)_{\omega_0^{(\alpha)}}^{-1}\mathbf{s}^{(\alpha)}.
\end{equation}
where $\alpha=\{i,r,t\}$ and
$k_0^{(\alpha)}$=$k{(\omega_0^{(\alpha)})}$ is the amplitude of the
wave vector corresponding to the frequency $\omega_0^{(\alpha)}$ at
which the spectrum $E^{(\alpha)}(\omega)$ has a maximum in the
amplitude. This definition is due to the well known
stationary phase method \cite{Brillouin}. As we are
interested in highlighting the effect of Eq.\ref{conserv energia} on
group velocities, it is necessary to make some further hypotheses
about the spectrum of the incident field as well as on the behavior
of $T(\omega)$ and $R(\omega)$. In particular, to exclude
interactions producing very distorded wave packets, for which the definition of group velocity
would be questionable, we
will assume that the spectrum of the incident pulse is narrower than
the transmission or reflection bands. This guarantees that the
spectral range of the incident pulse is sufficiently limited to include
at most one maximum (minimum) of $T(\omega)$ ($R(\omega)$) or
vice versa. However, also under such an hypothesis there is some
pathological case that must be excluded. This happens when the
frequency spectrum of the incident pulse has its maximum exactly
at the point where $T(\omega)=1$ (or $R(\omega)=1$). In
such a case the reflected (transmitted) pulse would be, in practice,
either completely absent or very distorted. Generally speaking, if the modulus of the incident
field $|E^{(i)}(\omega)|$ has a maximum (for the intensity of the beams we are obviously interested only in moduli) for a given value of
$\omega$ (determinated by the condition $d
|E^{(i)}(\omega)|/d\omega=0$), which we denote as $\omega_0^{(i)}$,
the reflected and transmitted spectra will show a shift in
their maxima. They will be respectively determined by the following
two equations
\begin{eqnarray}\label{spettrotx}
\frac{d |E^{(i)}(\omega)|}{d\omega}\left|t(\omega)\right|+\frac{d
\left|t(\omega)\right|}{d\omega}|E^{(i)}(\omega)|
=0\\\label{spettrorx} \frac{d
|E^{(i)}(\omega)|}{d\omega}\left|r(\omega)\right|+\frac{d
\left|r(\omega)\right|}{d\omega}|E^{(i)}(\omega)| =0
\end{eqnarray}
All the three fields would have a maximum in
$\omega_0^{(i)}$ if, and only if, in that point there is a maximum
(or a minimum) for $T(\omega)$ or $R(\omega)$ \cite{Nota1}. In such a case the fields would have the same group
velocity. In all other situations, i.e. when
$dT(\omega)/d\omega|_{\omega_0^{(i)}}=-dR(\omega)/d\omega|_{\omega_0^{(i)}}\neq0$
(where $\omega_0^{(t)}$ and $\omega_0^{(r)}$ denote the solutions of Eqs.
\ref{spettrotx} and \ref{spettrorx} respectively), we can evaluate
the moduli for the new group velocities as
\begin{equation}\label{vgtx}
v_g^{(t)}=\left(\frac{d\omega}{dk}\right)_{k_0^{(t)}}=v_g^{(i)}
+ \left(\frac{d^2\omega}{dk^2}\right)_{k_0^{(i)}} (k_0^{(t)}-k_0^{(i)})
\end{equation}
\begin{equation}\label{vgrx}
v_g^{(r)}=\left(\frac{d\omega}{dk}\right)_{k_0^{(r)}}=v_g^{(i)}
+ \left(\frac{d^2\omega}{dk^2}\right)_{k_0^{(i)}} (k_0^{(r)}-k_0^{(i)})
\end{equation}
where a Taylor expansion has been made. In Eqq. \ref{vgtx} and
\ref{vgrx} we have obviously $k_0^{(t)}\neq k_0^{(r)}$. The
reflected and transmitted pulses {\it cannot} have same group
velocities. This property originates from the observation that, by
virtue of the constraint described by the Eq.\ref{conserv energia},
spectral changes shown by the two emerging fields are not
independent. In fact, it is easily seen that every time
$T(\omega)$ grows, in an interval of $\omega$, then
in the same interval $R(\omega)$ must necessarily decrease and vice
versa. This implies that they must always have different velocities,
both different from the group velocity of the incident pulse.
{\it The difference between group velocities reported in Eqq.
\ref{vgtx}-\ref{vgrx}, depends on the dispersive properties of the
propagation environment (the second derivative term) and on the
properties of the optical structure considered (the $k$'s difference
term). } Of course in any non-dispersive
environment, i.e. when the modulus of wave vector is given by
$k=n\omega/c$ with $n$ constant (for instance $n=1$ for
the vacuum), the second derivative terms in Eqs.
\ref{vgtx}-\ref{vgrx} are null. Therefore in such a case the above
discussed phenomenon is absent. \\The competing effect of the group
velocity dispersion ($\Delta v_g$) must also be taken into account:
\begin{equation}\label{gvd}
\Delta v_g=\left(\frac{d^2\omega}{dk^2}\right)_{k_0} \Delta k.
\end{equation}
It holds for the spread of each single wave packet due to the effect
of the dispersive medium. This phenomenon tends to destroy the
packet by spreading it while propagating \cite{Berkley}. Hence only
structures where this effect is under control will be considered.\\
As an illustration let us calculate the difference for the group
velocities of the two emerging pulses in air after an interaction
with a dielectric slab. We can evaluate the second derivative term
in Eqs. \ref{vgtx},\ref{vgrx} with an accuracy of 1 ppm \cite{Nota2}
by using the most accurate recent measurement of the refractive
index of air \cite{refractAir} which improves Cauchy's formula for
the same refractive index \cite{BornWolf},
$n(\omega)=1+A(1+B\omega^2)$ with $A=28.79\times10^{-5}$ and
$B=1.6\times10^{-33}s^2$. As to the optical structure, we consider a
homogeneous, non-dispersive, rectangular dielectric slab with
refraction index $n=1.5$ and thickness $d=0.3 \mu m$ along the $x$
axis. The incident field impinging on this structure is supposed to
be a broad-spectrum gaussian light pulse (white light), centrered
around $0.555 \mu m$:
$E^{(i)}(\omega)=E^{(i)}_0\exp\{-(\omega-\omega_0^{(i)})^2/2\sigma_0^2\}$,
with $\omega_0^{(i)}=3.39397\times10^{15} rad/s$ and
$\sigma_0=4\times10^{14} rad/s$. It is well known that, in this
case, the reflectivity $\mathcal{R}_{\bot}$ for the interface
between air/glass depends on the polarization features of the field
under study as well as on the (mean) angle of incidence. For the
case of orthogonal polarization (electric field perpendicular to the
$x,y$ plane)
$\mathcal{R}_{\bot}=\sin^2(\theta_i-\theta_t)/\sin^2(\theta_i+\theta_t)$.
Here $\theta_i$ and $\theta_t$ represent the incident and
transmitted angle, respectively. For $\theta_i=65^\circ$ it follows
from Snell's law, $\theta_t\cong 37^{\circ}$ and
$\mathcal{R}_\bot\cong 0.228$. The transmission properties of the
entire optical structure are given by
\begin{equation}
T(\omega)=\frac{1}{1+F \sin^2(\alpha)}
\end{equation}
where $F=4\mathcal{R}_{\bot}/(1-\mathcal{R}_{\bot})^2$,
$\alpha=\omega n d \cos (\theta_t)/c$ \cite{BornWolf}. In
Fig.\ref{spettri} (subplot a) we show the frequency spectrum of the
incident pulse together with the transmission and reflection spectra
of the dielectric slab (which is, after all, a Fabry-Perot multiple
beam resonator), while in subplot b we show the frequency spectra
for incident, transmitted and reflected fields \cite{Risultati}. It
is manifest that the transmitted and reflected pulses are shifted,
with respect to the incident one, but in two opposite directions.
Indeed we obtain:
\begin{center} $v_g^{(t)}-v_g^{(i)}=2.47\times10^2 m/s$ \end{center}
\begin{center} $v_g^{(r)}-v_g^{(i)}=-1.65\times10^2 m/s$ \end{center}Recalling that the speed of light is known to an accuracy of $\pm 1 m/s$, it is evident that such an effect is
observable.\\
Summarizing, we investigated the role played by energy conservation on the
group velocity of a light pulse, propagating in dispersive media,
after an interaction with a generic linear, non-absorbing and
dispersive structure. We have shown that in general, transmitted and
reflected fields emerge with different group velocities. In a
forthcoming paper the extension of the present formalism to the case
of absorptive media will be considered.

\clearpage

\clearpage

\clearpage


\begin{thebibliography}{999}
\bibitem{Nota1}
This condition requires $dT(\omega)/d\omega|_{\omega_0^{(i)}}=-dR(\omega)/d\omega|_{\omega_0^{(i)}}=0$ and
$d^2T(\omega)/d\omega^2|_{\omega_0^{(i)}}=-d^2R(\omega)/d\omega^2|_{\omega_0^{(i)}}\neq
0$, in order to exclude cases of horizontal flex points. In addition further requirements on the second order derivatives must be fulfilled to assure that $\omega_0^{(i)}$ is a maximum for both fields, and not for only one of them. For instance, if in $\omega_0^{(i)}$ there is a maximum of $T(\omega)$, then there will be also a maximum for the reflected field only if it holds $\left|\frac{d^2|E^{(i)}(\omega_0^{(i)})|^2}{d\omega^2}\right|_{\omega_0^{(i)}}R(\omega_0^{(i)})>|E^{(i)}(\omega_0^{(i)})|^2\left|\frac{d^2R(\omega)}{d\omega^2}\right|_{\omega_0^{(i)}}$
\bibitem{Nota2}
An evaluation of $d^2\omega(k)/dk^2$ in eqq.\ref{vgtx}-\ref{vgrx}
does not require the knowledge of the function $\omega(k)$ i.e. the
inverse function of the dispersion relation
$k(\omega)=n(\omega)\omega/c$. In fact, it is possible to note that
if $k(\omega)$ is invertible around a frequency $\tilde\omega$ such
that $\tilde k=k(\tilde\omega)$, then, after some algebra, we have
\begin{equation}\label{d2}
\left(\frac{d^2\omega}{dk^2}\right)_{\tilde
k}=-\left(\frac{d^2k}{d\omega^2}\right)_{\tilde\omega}\left[\left(\frac{dk}{d\omega}\right)_{\tilde\omega}\right]^{-3}
\end{equation}
\bibitem{Risultati}
In the cited example, summarized by fig.\ref{spettri}, we have:\\
$\omega_0^{(t)}=3,3061\times10^{15}rad/s$,\\
$\omega_0^{(r)}=3,4527\times10^{15}rad/s$,\\
$d^2\omega/dk^2|_{k_0^{(i)}}=-8,42\times10^{-4}m^2/s$ and\\
$v_g^{(i)}=299.701.445 m/s$.
\end{thebibliography}
\end{document}